\documentstyle[11pt,epsfig]{article}
\def\edth{\;\raise1.0pt\hbox{$'$}\hskip-6pt\partial\;}
\def\baredth{\;\overline{\raise1.0pt\hbox{$'$}\hskip-6pt
\partial}\;}
\def\gsim{~\rlap{$>$}{\lower 1.0ex\hbox{$\sim$}}}
\def\lsim{~\rlap{$<$}{\lower 1.0ex\hbox{$\sim$}}}

\def\PRL{Phys. Rev. Lett.}
\def\PRD{Phys. Rev. D}
\def\MNRAS{Mon. Not. R. Astron. Soc.}
\def\apj{ ApJ}

\def\be{\begin{equation}}
\def\ee{\end{equation}}
\def\bea{\begin{eqnarray}}
\def\eea{\end{eqnarray}}

\begin{document}

\title{Improved Dark Energy Detection through the
  Polarization-assisted WMAP-NVSS ISW Correlation} 

\author{Guo-Chin, Liu$^{1}$, Kin-Wang, Ng $^{2,3}$,
        and Ue-Li, Pen$^{4}$}
\maketitle

$^{1}${\it Department of Physics, Tamkang University, 251-37
  Tamsui, Taipei County, Taiwan 251, R.O.C.}

$^{2}${\it Institute of Astronomy and Astrophysics, Academia
Sinica, Taipei, Taiwan 115, R.O.C.}

$^{3}${\it Institute of
Physics, Academia Sinica, Taipei, Taiwan 115, R.O.C.}

$^{4}${\it Canadian Institute for Theoretical Astrophysics,
University of Toronto, 60 St. George Street, Toronto, ON M5S 3H8,
Canada}

\begin{abstract}
Integrated Sachs-Wolfe (ISW) effect can be estimated by
cross-correlating Cosmic Microwave Background (CMB) sky with
tracers of the local matter distribution. At late cosmic time, the
dark energy induced decay of gravitation potential generates a
cross-correlation
signal on large angular scales.  The dominant noise
are the intrinsic CMB anisotropies from the inflationary epoch.
In this {\em Letter} we use CMB polarization to reduce this
intrinsic noise. We cross-correlate the microwave
sky observed by Wilkinson Microwave Anisotropy Probe (WMAP) with
the radio source catalog compiled by NRAO VLA Sky Survey (NVSS) to
study the efficiency of the noise suppression . We find that the
error bars are reduced about $5-12\%$, improving the statistical power.
\end{abstract}


Recently released data made by Wilkinson Microwave Background
Probe (WMAP)~\cite{jaroski} is  used to study different
cosmological models with an unprecedented accuracy. The observed
data puts tight constraints on intrinsic properties of the
Universe such as the geometry, the matter content, the origin of
inhomogeneities, and the formation of large scale structures.
Furthermore, combining observations from type Ia
supernovae~\cite{knop} and large scale structures (e.g. Allen et
al. 2002~\cite{allen}) converges our understanding of the Universe to a
concordance model. In this model, the present Universe is
dominated by dark energy with negative pressure which is
accelerating at late times. Although dark energy can
explain the acceleration, we  know very little about its
nature and origin. More observations are definitely necessary for
constraining this dark energy component.

As CMB photons propagate through a gravitational potential generated
by large scale structures in the expanding Universe, the photons
undergo an energy shift. In a matter-dominated universe, the
gravitational potential stays a constant, so the CMB temperature
fluctuations generated through this so-called Sachs-Wolfe (SW)
effect~\cite{SWE} depend only on the potential difference between the
recombination epoch and the present time.  However, the existence of a
spatial curvature or dark energy results in a time-varying
gravitational potential, thus producing, in addition to the SW effect,
time-integrated temperature fluctuations of the CMB. This is known as
the integrated Sachs-Wolfe (ISW) effect~\cite{SWE}. Therefore, the
detection of the late-time ISW effect in a flat universe can be
regarded as direct dynamical evidence for dark energy. Furthermore,
measuring the ISW effect provides an independent method to probe the
properties of dark energy.

The ISW effect manifests itself on large angular scales because the
CMB temperature fluctuations or anisotropies are dominantly induced by
the gravitational potential at large scales. On these angular scales,
the CMB signal is dominated by the primary anisotropies generated from
the recombination epoch at redshift $z \simeq 1100$ as well as from
the reionization epoch at redshift $z \simeq 10$. It is not easy to
isolate the ISW effect from the primary CMB fluctuations.  The
precision of large-scale anisotropy power measurement is limited by
the unavoidable cosmic variance. Therefore, attempts in measuring the
ISW effect solely from CMB experiments may not be able to give tight
constraints on dark energy models.

However, a positive large-scale correlation signal may occur by
correlating the CMB sky with the local matter distribution as a
result of the ISW effect. This idea was first explored by
Crittenden and Turok (1996). Several authors tried to detect the
ISW effect by correlating CMB satellite data with different matter
tracers.  The CMB anisotropy data made by the COsmic Background
Explorer (COBE) was firstly used for this
study~\cite{Boughn-Crittenden}. But the authors concluded that the
resolution and sensitivity of COBE were not good enough for a
detection. Recently, the WMAP mission has provided a set of
high-quality CMB data. As to the matter distribution, several
tracers have been used for this study, such as radio sources
provided by NRAO VLA Sky Survey (NVSS)~\cite{nvss}, hard X-ray
data from High Energy Astronomy Observatory-1 satellite
(HEAO-1)~\cite{heao}, Sloan Digital Sky Survey (SDSS)
data~\cite{sdss}, and Two Micron All Sky Survey Extended Source
Catalog (2MASS XSC)~\cite{2mass}. Douspis et al. 2008~\cite{douspis}
has investigated the properties of a survey required to detect this
correlated signal. The data analyses of the
cross-correlation are performed in real~\cite{Boughn-Crittenden,
bc2004,nolta}, harmonic~\cite{afshordi}, and wavelet~\cite{vielva,
mcewen} spaces.

A significant uncertainty in the cross-correlation is coming from
the spurious correlation of matter distribution with the primary
CMB anisotropies generated from the recombination and reionization
epochs. It plays a role like the cosmic variance in observations
of the low multipoles of the CMB anisotropy that have only a few
independent modes. This spurious correlation obscures the
measurement of the true correlation that we are interested in and
indeed weakens the constraint on the dark energy component.
However, CMB polarization is also generated when the primary
anisotropies are scattered by free electrons in these two epochs.
Using the fact that the ISW effect occurs at relatively late
times, the information imprinted on the CMB polarization may give
an opportunity to separate the ISW effect from the primary
anisotropies. If this can be done, we will suppress or even get
rid of the spurious correlation. In this {\em Letter}, we
construct a simple relation between the CMB polarization and its
corresponding primary anisotropies. This relation is found to be
weakly dependent on the emerging dark energy at late times, thus
allowing us to subtract the primary CMB anisotropies from the CMB
anisotropy data to obtain the genuine ISW signal. Then, the
resulting anisotropies are used to correlate with matter tracers
to obtain a better signal-to-noise detection of the
cross-correlation. 



Instead of using Stokes parameters $Q$ and
$U$~\cite{chandrasekar1960}, CMB polarization is conventionally
characterized by a divergence free component, commonly called the  $E$
mode, and a curl component, the  $B$ mode. Small-scale
polarization of the CMB is generated through Thomson scattering of
the CMB anisotropic radiation by free electrons at the
recombination epoch. For the angular scales at which the
cross-correlated ISW effect is concerned,  $E$-mode polarization
is generated by re-scattering off the free electrons at
reionization epoch, and $B$-mode polarization is also generated in
the presence of tensor-mode perturbations, predicted for example
in inflation model. So far, there is no evidence for the $B$ mode
polarization, so we will not consider it.

Since CMB polarization is generated from temperature anisotropies,
we expect some relation between the $E$-mode polarization and the
CMB anisotropies. We simply assume that the relation can be
written as
\begin{equation}
E_{\ell}^{\rm NOISW}=a_{\ell}T_{\ell}^{\rm NOISW}+n_{\ell}.
\label{ETrelation}
\end{equation}
Here $E_{\ell}^{\rm NOISW}$ and $T_{\ell}^{\rm NOISW}$ are the
$E$-mode polarization and temperature anisotropies in each
multipole $\ell$, each coefficient $a_{\ell}$ quantifies the
amount of the correlation between the $E$-mode polarization and
the temperature anisotropies, and $n_{\ell}$ plays a role like the
``noise'' of this relation. For our purpose, we construct this
relation in the situation in which the ISW effect is removed.

We define the ``$E$-mode polarization corrected '' temperature
anisotropies as
\begin{equation}
\tilde{T}^{\rm E-COR}_\ell =  T_\ell - E_\ell/a_\ell \simeq T_\ell
- E^{\rm NOISW}_\ell/ a_\ell. \label{estimate}
\end{equation}
We have made the approximation in the last equality due to the
fact~\cite{ng,cooray-melchiorri} that the polarization contributed
by the ISW effect is significantly smaller than that by the
primary temperature quadrupole re-scattering at the reionization.
Then, we cross-correlate $E_{\ell}^{\rm NOISW}$ in
equation~(\ref{ETrelation}) with the CMB temperature anisotropies:
\begin{equation}
\langle TE\rangle_\ell^{\rm NOISW}=a_\ell \langle
TT\rangle_\ell^{\rm NOISW}, \label{TE}
\end{equation}
and obtain the auto-correlation as
\begin{equation}
\langle EE\rangle_\ell^{\rm NOISW}=a_\ell^2 \langle
TT\rangle_\ell^{\rm NOISW}+ n_\ell^2. \label{EE}
\end{equation}
Thus, we can calculate the sets $a_\ell$ and $n_\ell$ using
equations~(\ref{TE}) and (\ref{EE}), given the power spectra
$C_{T\ell}$, $C_{C\ell}$, and $C_{E\ell}$ from the theoretical
prediction. Once the two sets are determined, they can be applied
to real data. However, in numerical practice, we have found that
the numbers of $a_\ell$ are quite small and the noise term
$n_\ell^2$ is in fact comparable to $\langle EE\rangle_\ell$. It
means that $T_\ell-E_\ell/a_\ell$ may result in an over
subtraction due to the noise $n_\ell$. To circumvent this problem,
we multiply each $E_\ell/a_\ell$ by a Wiener filter constructed as
\begin{equation}
W_{1\ell}=\langle TT\rangle_\ell/(\langle
TT\rangle_\ell+n_\ell^2/a_\ell^2).
\end{equation}

In Fig.~1, we plot the $E$-mode polarization corrected power
spectrum (denoted by the dot-dashed curve) of the temperature
anisotropies, $\langle\tilde{T}\tilde{T}\rangle^{\rm E-COR}_\ell$,
where $\tilde{T}^{\rm E-COR}_\ell$ is defined in
equation~(\ref{estimate}). In determining $a_\ell$ and $n_\ell$,
we have generated $500$ CMB sky maps, using the power spectra
obtained from running the CMBFast code~\cite{CMBFast} with the
WMAP best-fit cosmological parameters. For the necessary power
spectra used in equations~(\ref{TE}) and (\ref{EE}), we have
removed the ISW effect in the CMBFast code by forcing the
time-varying gravitational potential to vanish for redshift $z<5$.
For the comparison, we also plot the total power spectrum of the
temperature anisotropies and the isolated ISW effect, respectively
denoted by the solid and the dashed curves. We have found that we
can filter out more than about $10\%$ of the primary temperature
anisotropies for the multipoles $\ell < 30$. For $40 < \ell < 70$,
the correction is inefficient due to the low $\langle TE\rangle$
correlation from the theoretical prediction. More efficient
correction occurs for $ 80 <\ell< 100$ (in Fig.~1); however, the
cross-correlation signal of the CMB and large-scale-structure is
small at this angular scale~\cite{CT}.

\begin{figure}[htbp]
\centerline{\psfig{file=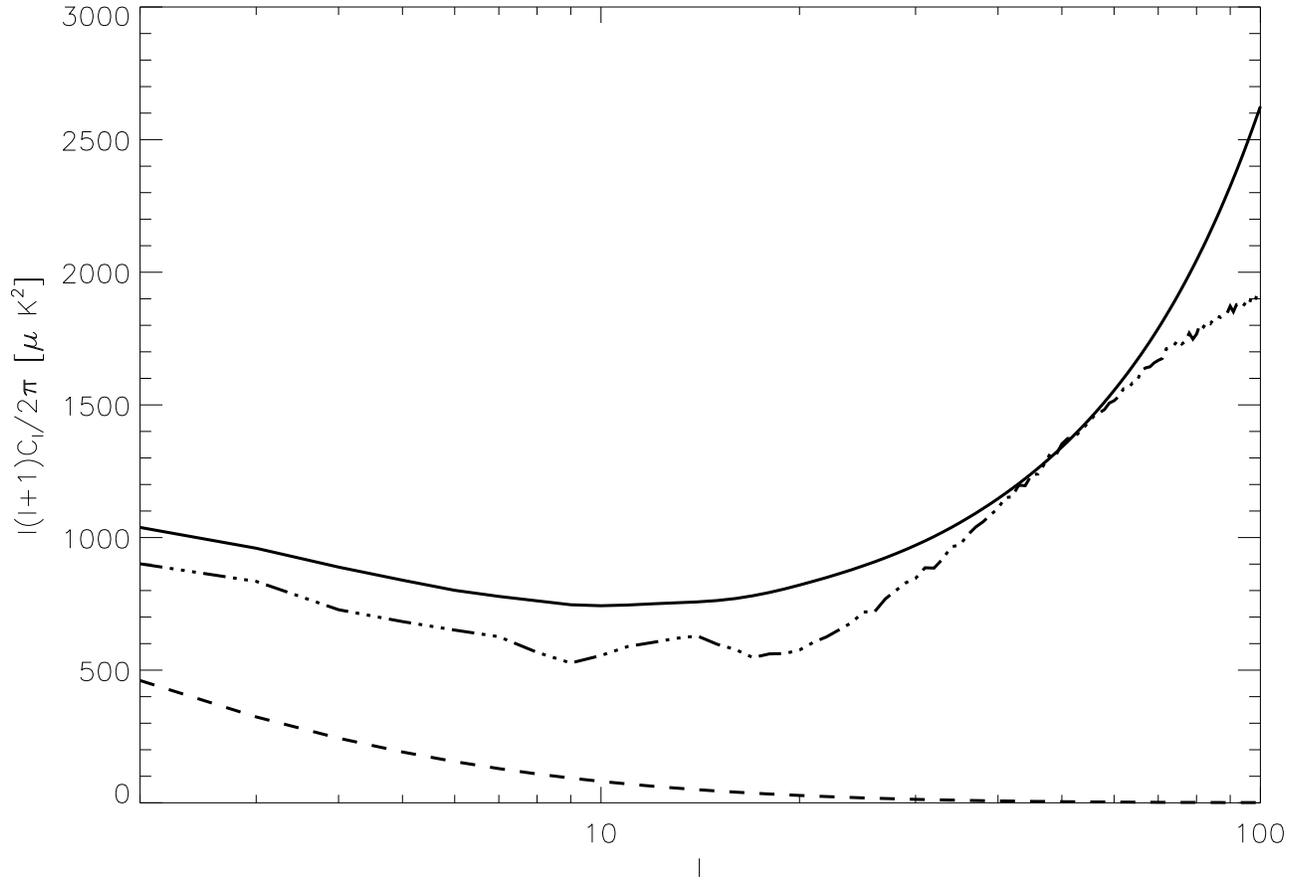}} \caption{Power spectrum of the
CMB temperature anisotropies after the correction by the $E$-mode
polarization (dot-dashed curve). For comparison, we also plot the
total power spectrum before correction (solid curve) and the ISW
effect (dashed curve). Using information of $\langle TE\rangle$
correlation, we can filter out the part of the signal coming from
redshifts before $z=5$. }
\label{CMB_ISW}
\end{figure}


Now we apply the ``corrected'' temperature anisotropies described
above to calculate its cross-correlation with the matter
tracers. We use the WMAP 7-year foreground reduced maps with full
resolution. We take only the Q and V frequency channels because of
the low contamination of the foreground emission on both
temperature anisotropies and polarization~\cite{jaroski}. For
the matter distribution, we use NVSS radio sources. NVSS is
operated at $1.4$ GHz with flux limit $2.5$ mJy. It is complete
for declination $\delta > -40^o$ and contains 1.8 million sources.
Its $82\%$ sky coverage provides good information at large angular
scales. Though the redshifts of individual radio sources are
largely unknown, the luminosity function~\cite{dunlop} indicates
that they are distributed in redshift range $0<z<2$, with a peak
distribution at $z \sim 0.8$. Crittenden and Turok (1996) have
studied the signal contribution of the cross-correlation with
different cutoffs in redshift~\cite{CT} and concluded that the
redshift distribution of the NVSS radio sources is very suitable
for this study. The cross-correlation power spectrum between NVSS
and WMAP is simply calculated by
\begin{equation}
C^{NW}_{\ell}=\frac{1}{2\ell+1}\sum_{m} \langle a^N_{\ell
m}a^{*W}_{\ell m}\rangle,
\end{equation}
where $a^N_{\ell m}$ and $a^{W}_{\ell m}$ are the harmonic
coefficients of the NVSS and WMAP maps, respectively. We
distribute the NVSS catalog in the Healpix scheme~\cite{healpix}
at resolution-9 map, whose resolution is the same with the WMAP
7-year foreground reduced maps. Such map has $12{N}_{side}$ pixels
with the same area in all sky, where $N_{side}=2^9$. We calculate
the coefficients of the spherical harmonics $a_{lm}$ for the
temperature anisotropies, the $E$ mode polarization, as well as
the NVSS sources, using the Healpix package with a mask which is
obtained by combining the standard polarization mask (P06) used by
the WMAP team and the blank sky of the NVSS survey. Applying this
mask gives a sky coverage of about $56\%$ and the average number
density of NVSS radio galaxies is $171494/sr$.


In Fig.~2, we plot the signal of the cross-correlation of WMAP and
NVSS for the Q and V frequency bands. To reduce the correlation
among neighboring multipoles, we bin our results into five bins
with equal spaces in logarithm-$\ell$ from $2$ to $100$ (We found
that masking about $44\%$ of the sky would cause about $44\%$
correlation among neighboring multipoles). The instrumental noise
in the WMAP data at such large angular scales is negligible when
compared to temperature anisotropies, but it is comparable to the
$E$ polarization signal. Therefore, in practice, we find the {\em
estimated} ISW temperature as
\begin{equation}
\tilde{T}^{\rm
E-COR}_{\ell}=T_\ell-W_{1\ell}W_{2\ell}E^{\rm
    NOISW}_\ell/a_\ell,
\end{equation}
with one more Wiener filter given by
\begin{equation}
W_{2\ell}=\langle EE\rangle_\ell/(\langle EE\rangle_\ell+N_\ell),
\end{equation}
where $N$ is the covariance matrix for the instrumental noise. We
have calculated $N$ by the thermal noise in each pixel provided by
WMAP team without taking into account the correlated noise between
different pixels. It may cause the wrong power estimation in
low-$\ell$ multipoles. We also need to notice that we cannot
obtain the exact values of the multipoles by this method due to
the incomplete coverage of the sky. In case of modelling or
parameter fitting, we need to estimate these two factors
carefully.

The spurious correlation resulted from the primary CMB temperature
anisotropies can be quantified using simulated, uncorrelated CMB
maps. We use the $500$ simulated CMB sky maps in the previous
section to do the blind correlation with the real NVSS data. The
error bars in Fig.~2 show the $1\sigma$ region derived from this
result. Let us define $\chi^2=\sum_i
(\hat{C}_{Bi}-C_{Bi})^2/\sigma^2_{Bi}$, where $\hat{C}_{Bi}$ and
$C_{Bi}$ are respectively the measured and predicted band powers
of the WMAP-NVSS cross-correlation, and $\sigma_{Bi}$ is the
$1\sigma$ error of the $i$th band. For a model with null
correlation, i.e., $C_{Bi}=0$, we find that $\chi^2=9.8$ for the
Q band and $\chi^2=9.0$ for the V band in the case without the
$E$-polarzation correction. After the $E$-polarization correction,
the values increase to $\chi^2=11.7$ for the Q band and
$\chi^2=10.6$ for the V band. We also find that the correction
improves the errors resulted from the spurious correlation by
$6.0\%$, $9.5\%$, $12.3\%$, $4.9\%$, and $7.3\%$ in the five respective
bins.

We acknowledge the extensive use we have made of data from the
WMAP satellite, and thank Dr.~L.~Chiang for fruitful discussions on
the WMAP data. We also appreciate discussions with Drs.~N.~Aghanim and
M.~Douspis. G.-C. Liu (K.-W.~Ng) thanks for the
support from the National Science Council of Taiwan under the grant
NSC97-2112-M-032-007-MY3 (NSC98-2112-M-001-009-MY3) and
support from the Institute of Physics, Academia Sinica.

\begin{figure}[htbp]
\centerline{\psfig{file=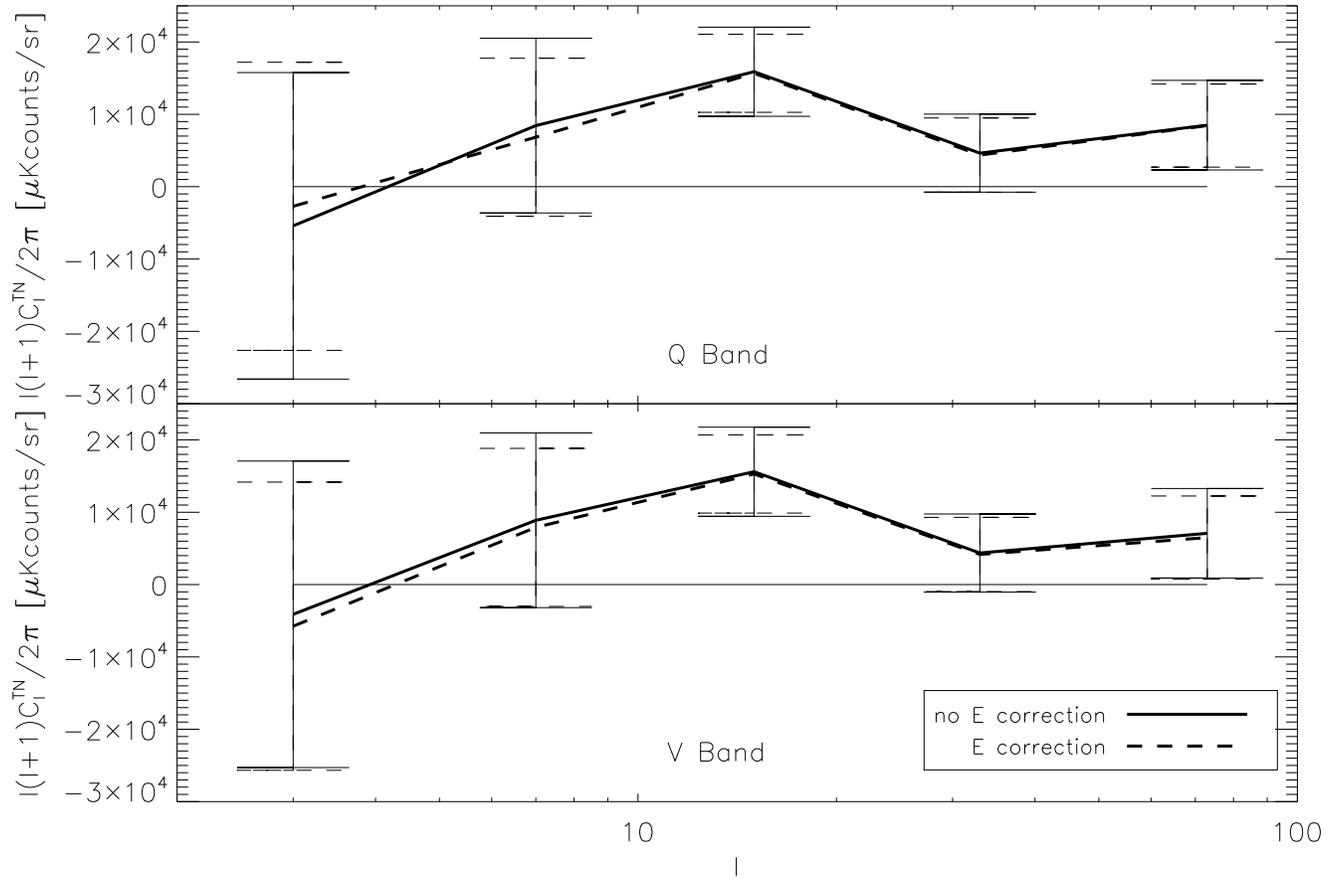}}
\caption{Cross-correlation of WMAP and NVSS for Q and V frequency
  bands. The erros bars are calculated by cross-correlated $500$
  simulated CMB skys with NVSS catalog. Using $E$-polarization, the
  error bars and signal of cross-correlation are changed.
}
\label{phase}
\end{figure}

\end{document}